\documentclass[aps,pre,superscriptaddress,twocolumn,amsmath,amssymb]{revtex4}
\usepackage[dvips]{graphicx}
\usepackage[usenames,dvipsnames]{color}
\usepackage{newcent}  
\usepackage{natbib}
\newcommand{\BE}{\begin{equation}}
\newcommand{\EE}{\end{equation}}
\newcommand{\BA}{\begin{eqnarray}}
\newcommand{\EA}{\end{eqnarray}}
\newcommand{\br}{\bf r}
\newcommand{\bx}{\bf x}
\newcommand{\bk}{\bf k}
\newcommand{\bs}{\bf s}

\begin{document}

\title{Minimal mechanisms for vegetation patterns in semiarid regions}
\author{Ricardo Mart\'inez-Garc\'ia}
\affiliation{IFISC, Instituto de F\'isica Interdisciplinar y Sistemas Complejos (CSIC-UIB),
 E-07122 Palma de Mallorca, Spain}

\author{Justin M. Calabrese}
\affiliation{Conservation Ecology Center, Smithsonian Conservation Biology Institute, National Zoological Park,
1500 Remount Rd., Front Royal, VA 22630, USA.} 

\author{Emilio Hernandez Garcia} 
\affiliation{IFISC, Instituto de F\'isica Interdisciplinar y Sistemas Complejos (CSIC-UIB),
 E-07122 Palma de Mallorca, Spain}

\author{Crist\'obal L\'opez} 
\affiliation{IFISC, Instituto de F\'isica Interdisciplinar y Sistemas Complejos (CSIC-UIB),
 E-07122 Palma de Mallorca, Spain}


\begin{abstract}

The minimal ecological requirements for formation of regular
vegetation patterns in semiarid systems have been recently
questioned. Against the general belief that a combination of
facilitative and competitive interactions is necessary, recent
theoretical studies suggest that, under broad conditions,
nonlocal competition among plants alone may induce patterns. In
this paper, we review results along this line, presenting a
series of models that yield spatial patterns when finite-range
competition is the only driving force. A preliminary derivation
of this type of model from a more detailed one that considers
water-biomass dynamics is also presented.
Keywords:
Vegetation patterns, nonlocal interactions
\end{abstract}

\maketitle

\section{Introduction}

Vegetation in semiarid regions around the world can
form striking, highly organized patterns.
Many approaches have been used to tackle the
study of vegetation patterns both from a theoretical
and a empirical side. Many works have focused
on measuring the different types of interactions
among plants that are present in water-limited
 systems as well as their spatial
 ranges and strength \citep{Dunkerley,Barbier2008}.
On the theoretical side, which is the focus
 of this paper, mathematical models have been proposed either accouting for
the evolution of the vegetation biomass alone \citep{Martinez-Garcia2013,Lefever1997,D'Odoricoa,Lefever2012}
or coupled with the dynamics of the water
 in the system \citep{VonHardenberg2001,Gilad2004}.
A common point of all these studies is the view of
 the pattern formation phenomenon as a
symmetry-breaking process that induces an
instability on the uniform vegetation state
 \citep{Klausmeier1999,Lefever1997,Lejeune1999}.

Interest in plant patterns stems from the idea that
these structures provide information about the physical and
biological processes that generate them. However, the same
strength of the modern approach to vegetation patterns, that
is, its {\it universality}, becomes a great
disadvantage when searching for relationships between patterns
and processes, as many different processes can give rise to the
same spatial structures. As a result, it is useful on the
theoretical side to unveil the minimal set of biophysical
mechanisms under which typically-observed patterns may appear
in water-limited systems. Most existing mathematical models of
vegetation pattern formation assume an interplay between
short-range facilitation and long-range competition. While it
is clear that such a combination of mechanisms is likely
responsible for patterns in some conditions---for example
regular stripes on hillsides \citep{Klausmeier1999}---whether
or not both mechanisms must always be present for pattern
formation is an open question. While competition for water is
likely the key factor for semiarid systems, some studies
\citep{Rietkerk2008, Martinez-Garcia2013a} have suggested that
local facilitative interactions maybe unnecessary, or of only
minor importance, for pattern formation. Following these ideas,
the authors have recently introduced a model of vegetation
density for water-limited regions where only competition among
plants is considered \citep{Martinez-Garcia2013}. Here the
interaction enters by allowing the growth rate of a plant to
diminish with the number of other individuals competing with it
for resources (water). Despite the fact that facilitation is
ignored, this non-local competition model produces a
spectrum of spatial patterns similar to the one observed in
models assuming both facilitation and competition are
necessary.

In this paper we, extend the results of
\cite{Martinez-Garcia2013} to address several open questions:
1) Do patterns depend on how competition enters in the
dynamical equations? 2) What is the role of nonlinearities? 3)
Can simple models featuring nonlocal competition be derived
from  more fundamental ones that consider the dynamics of
plants and water sources? To answer these questions, we present
a set of nonlocal models with only competitive interactions
that enter in the equations either linearly or nonlinearly. In
the latter case, we complement our previous work by also
allowing nonlocal competition to enter in the death term.
Patterns emerge in all of these models, and in a
sequence related to the one observed in standard
facilitative-competitive models.
We also present preliminary results on how the
nonlocal density equations can be derived from a more
mechanistic dynamics that considers biomass and water
interactions.

More in detail, the outline of the paper is as follows. In
Section \ref{sec:revmodels} we give an overview of previous
nonlocal models and describe new ones: subsection
\ref{subsec:kernel} shows a review of standard kernel-based
descriptions with facilitative and competitive interactions; in
Subsection \ref{subsec:birth} we review the competition-only
model introduced in  \cite{Martinez-Garcia2013}; then in
Subsection \ref{subsec:death} we study the model where the
nonlocality enters in the death term; in Subsection
\ref{subsec:lineal} the model studied is of competition
entering linearly in the equations. In Section
\ref{sec:derivation} the derivation of density models from
water-biomass dynamics is discussed, and in Sec.
\ref{sec:conclusions} we write down our conclusions and
summary.

\section{Spatially nonlocal models for the tree-density}
\label{sec:revmodels}

Vegetation patterns arise from self-organization mechanisms due
to dynamic interactions among plants and between these and
their environmental conditions. Existing studies
\citep{Lejeune1999,Lefever1997,VonHardenberg2001,Klausmeier1999,
Rietkerk2002, Barbier2008,JGRG:JGRG103} consider two typical
length scales to account for facilitative (short-range) and
competitive (long-range) interactions. As mentioned, the need
for these two types of mechanisms has been recently questioned
in \cite{Martinez-Garcia2013} from a mathematical point of
view. In this section, we review the standard models which
include both facilitation and competition, and then present the
competition-only model of \cite{Martinez-Garcia2013}.

\subsection{Kernel-based models with facilitative and competitive mechanisms}
\label{subsec:kernel}

\begin{figure}
\noindent\includegraphics[width=0.4\columnwidth,angle=-90]{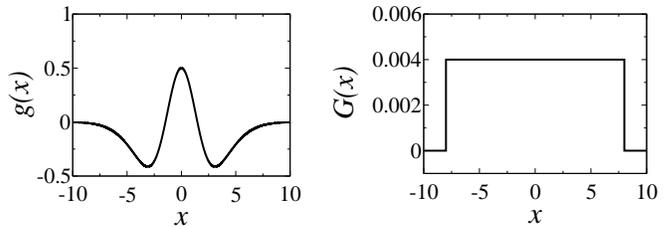}
\caption{(Left) Kernel function of standard one-dimensional
kernel-based models
considering both competitive and facilitative interactions.
It is built with a combination of positive and negative Gaussian functions,
$g(x)=1.5\exp\left(-(x/2)^{2}\right)-\exp\left(-(x/4)^{2}\right)$.
(Right) Competitive-only top-hat kernel with range $R=8$.
}
\label{fig0}
\end{figure}

The kernel-based models \citep{Borgogno2009} express vegetation
density mathematically as integro-differential equations with a
spatially nonlocal interaction function. Roughly
speaking, two types exist: a) those where the nonlocality
enters linearly (nonlinearities appear but without spatial
coupling), or b) those where the nonlocality enters
multiplicatively \citep{Lefever1997}. For simplicity here, we
only discuss the linear class, the so-called neural models
\citep{Murray2002}. The dynamics of the vegetation-density
field, $\rho (\br ,t)$, is given by:
\begin{equation}
\frac{\partial \rho}{\partial t}=F(\rho)+\int_{\Gamma} g(\br,\br')(\rho(\br')-\rho_0),
\label{neuralmodel}
\end{equation}
where $F(\rho)$ denotes the local dynamics whose steady state
is $\rho_0$, and $\Gamma$ is the spatial domain over which the
kernel function $g(\br, \br')$ is defined. The term
$\int_{\Gamma} g(\br,\br')\rho(\br')$ (assuming isotropy and
homogeneity it is more commonly expressed as $g(|\br-\br'|)$)
indicates that spatial interactions positively affect
(facilitation) the growth when $g>0$, and the contrary
(competition) when $g<0$. Interaction kernels in these models
typically exhibit the shape shown for the one-dimensional case
in  left panel of Fig. \ref{fig0}, and are thus positive at
short scales and negative at long-range. In fact, the way the
spatial structure emerges from Eq. (\ref{neuralmodel}) is easy
to understand: small perturbations larger than the homogenous
state, $\rho_0$, tend to increase locally due to the positive
interaction with nearby points, while those with $\rho<\rho_0$
decrease in the interaction neighborhood. Thus, short-range
facilitation enhances spatial heterogeneity and the long-range
inhibition (the negative part of the kernel) limits the
indefinite growth of the perturbation. A justification and
deeper analysis of these type of kernels for vegetation models
is given in \cite{Borgogno2009}. Biologically speaking, the
facilitation range is usually assumed to be similar to the
crown radius, while the competition range is related to the
lateral root length. While negative vegetation densities are
mathematically possible under these models, they are
biologically nonsensical. Therefore, works using kernel-based
models usually set negative densities to zero in numerical
simulations \citep{Borgogno2009}.

\subsection{A kernel-based model including only competitive interactions}
\label{subsec:birth}

Following previous studies \citep{Rietkerk2008,
Martinez-Garcia2013a} suggesting that vegetation patterns could
emerge without short-range facilitation, and assuming that
competition for water is the unavoidable interaction in arid
and semiarid systems, \cite{Martinez-Garcia2013} proposed a
nonlocal model with only competitive interactions. The equation
for vegetation density is
\begin{equation}\label{modelo}
\frac{\partial \rho({\bf x},t)}{\partial t}=\beta_0 r(\tilde{\rho},\delta)\rho({\bf x},t)(1-\rho({\bf x},t))-\alpha\rho({\bf x},t),
\end{equation}
where $\tilde \rho$ is the mean vegetation density within a
neighborhood, weighted with the kernel $G(x)$, around a given
spatial point:
\begin{equation}\label{nonlocaldensity}
\tilde{\rho}({\bf x},t)=\int G(|{\bf x}-{\bf x'}|)\rho({\bf x'},t)d{\bf x'}.
\end{equation}
The different terms in the model come from considering the
growth and death dynamics of vegetation. Population growth
follows a sequence of seed production, dispersal and
establishment:

\begin{enumerate}
\item
Production happens at rate $\beta_0$ per plant.
Assuming local seed dispersion and that all seeds
may give rise to new plants, the growth rate is $\beta_0 \rho$.
After a seed lands, it has to overcome competition to establish.
The two next competing mechanisms are taken into account:

\item
Space availability limits the density to a maximum value $\rho_{max}$,
so the proportion of available space at a point ${\bf x}$ is $1-\rho (\bx,t)/\rho_{max}$.
Density can be scaled such that $\rho_{max}=1$ and thus
the growth term is limited by a factor $(1-\rho (\bx,t))$.

\item
Once the seed has germinated, it competes with other plants for
water and other resources in the soil.
The probability of overcoming this competition is given
by $r(\tilde \rho, \delta)$. This function decreases when
$\tilde \rho$
increases, so that $r'(\rho_0,\delta)
 \equiv (\partial r /\partial\tilde \rho)_{\tilde\rho=\rho_0} <0$.
We assume that plants compete with other plants in their neighborhood,
which is defined by a distance of the order of twice the typical root length.

\end{enumerate}

It is worth stressing the difference between the function $G$
in this description and the $g$ in the previous subsection. $g$
contains the information about the interactions (cooperative
when positive and competitive when negative) present in the
system \citep{Lefever1997,D'Odorico2006}. Since these are of
facilitative and competitive type, the kernels are positive (at
short scales) and negative (at long scales). On the contrary,
$G$ is strictly positive and defines an influence region of a
focal plant which is used to compute an averaged density of
other plants around it. Also, nonlocal competition
enters nonlinearly, at variance with Eq. (\ref{neuralmodel}),
so that negative densities no longer appear.

Performing a linear stability analysis of the stationary solution,
$\rho_0$, of
Eq. (\ref{nonlocaldensity}) the perturbation growth rate is (see \cite{Martinez-Garcia2013})
for details
\begin{equation}
 \lambda({\bf k})=-\alpha\rho_{0}\left[\frac{1}{1-\rho_{0}}-
\frac{r'(\rho_{0},\delta)}{r(\rho_{0},\delta)}\hat{G}({\bf k})\right],
\label{relationdispersion}
\end{equation}
where $\hat{G}({\bf k})$ is the Fourier transform of the
kernel, $\hat{G}({\bf k})=\int G({\bf x})\exp(i{\bf k}\cdot{\bf
x})d{\bf x}$.

\begin{figure}
  \noindent\includegraphics[width=\columnwidth]{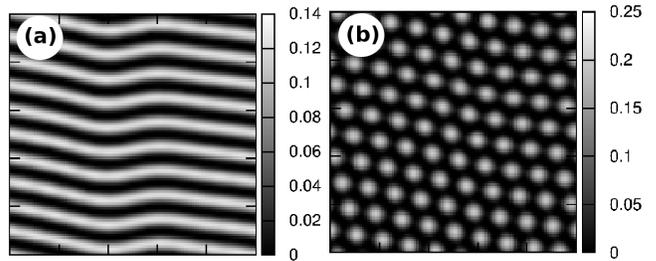}
  \caption{Close-to-stationary
 spatial structures shown by the model using the $r(\tilde{\rho},\delta)$ given by Eq.~(\ref{pldecay}).
  Darker grey levels represent smaller densities. (a) Vegetation stripes, $\delta=16.0$.
  (b) Vegetation spots, $\delta=17.0$.  Other parameters: $\beta_{0}=1.0$ and $\alpha=0.5$.}
  \label{fig1}
  \end{figure}

Since $r'<0$ equation (\ref{relationdispersion}) indicates that
patterns may appear ($\lambda >0$) in the model when $\hat G
(\bk)$ takes negative values, provided that competition is
strong enough. This may happen, for example, when the kernel
has a finite range (an example is shown in right panel of Fig.
\ref{fig0}), so that it is only different from zero (positive)
in a finite domain around $\bx =0$. In plant dynamics, this
finite range arises naturally from the length of the roots. The
model recovers the gapped and striped patterns observed in arid
and semiarid landscapes. Figure \ref{fig1} shows the stationary
patterns obtained by integrating Eq.~(\ref{model}) in a patch
of $10^{4}$~m$^{2}$ with periodic boundary conditions and a
competition range of $R=8$~m. $G$ is a two-dimensional top-hat
function (a cut across it will be similar to the right plot in
Fig. \ref{fig0})
 and the probability of overcoming nonlocal competition is given by
\begin{equation}\label{pldecay}
 r(\tilde{\rho},\delta)=\frac{1}{(1+\delta\tilde{\rho})},
\end{equation}
which makes$\rho_0$ analytically solvable. The patterns only appear if
the Fourier transform of the kernel function has negative
values. For the two-dimensional top hat kernel of width 2R, the
Fourier transform is $\hat G(k) = 2J_1(kR)/kR$, where $J_1$ is
the first order Bessel function \citep{hernandezpre2004}.

\subsection{Competition through a nonlocal nonlinear death term}
\label{subsec:death}

As a complement to the vegetation dynamics in
Eq.~(\ref{nonlocaldensity}) we next discuss a system, again
without facilitation, where resource competition enters through
the death rate. There is now a nonlocal nonlinear death
term resulting in a higher death rate when the surrounding
vegetation density increases. This is mathematically expressed
as:
 \begin{equation}\label{sav1}
\frac{\partial \rho({\bf x},t)}{\partial t}=\beta\rho({\bf x},t)(1-\rho({\bf x},t))-
\alpha_d \rho({\bf x},t),
\end{equation}
where $\alpha_d=\alpha_{0}h(\tilde{\rho}({\bf x},t),\delta)$ is
the nonlocal death rate ($\alpha_0$ is a constant and $h$ an
arbitrary function), and $\beta$ is the constant birth rate.
Nonlocal competition affecting mortality has been shown to
promote clustering in individual-based population models
\citep{Birch2006}.

As before, $\tilde{\rho}({\bf x},t)$ is the nonlocal density of
vegetation at the point ${\bf x}$, where $\tilde{\rho}({\bf
x},t)=\int\rho({\bf x'},t)G(|{\bf x}-{\bf x'}|)dx'$. $G$ is the
kernel function that defines an interaction range and modulates
its strength with the distance from the focal plant. Space
availability for a seed to establish appears in the birth term
via  $1-\rho({\bf x},t)$ (local competition).
$h(\tilde{\rho}({\bf x},t),\delta)$ gives the probability that
a plant dies as a function of competition for water with the
roots of other plants. Since it is a probability, $0<h<1$ and
it increases with increasing values of the averaged density,
$\bar \rho$, and the (positive) competition parameter,
$\delta$. The stationary solutions of Eq.~(\ref{sav1}),
$\rho_{0}$, are obtained by solving
 \begin{equation}
\beta\rho_{0}(1-\rho_{0})-\alpha_{0}h(\rho_{0},\delta)\rho_{0}=0,
 \end{equation}
which has a trivial solution, $\rho_{0}=0$ referring to the
bare-ground state, and a vegetated state that is obtained from
 \begin{equation}\label{vegsta}
 \beta(1-\rho_{0})-\alpha_{0}h(\rho_{0},\delta)=0,
  \end{equation}
once the function $h$ has been chosen.

A linear stability analysis of the stationary homogeneous
state, $\rho_{0}$, yields the dispersion relation
\begin{equation}
 \lambda({\bf k})=\beta(1-2\rho_{0})-\alpha h(\rho_{0},\delta)-\alpha\rho_{0}h'(\rho_{0},\delta)\hat{G}({\bf k}),
\end{equation}
where $\hat{G}({\bf k})$ is the Fourier transform of the kernel function.

\begin{figure}
\centering
\includegraphics[width=\columnwidth]{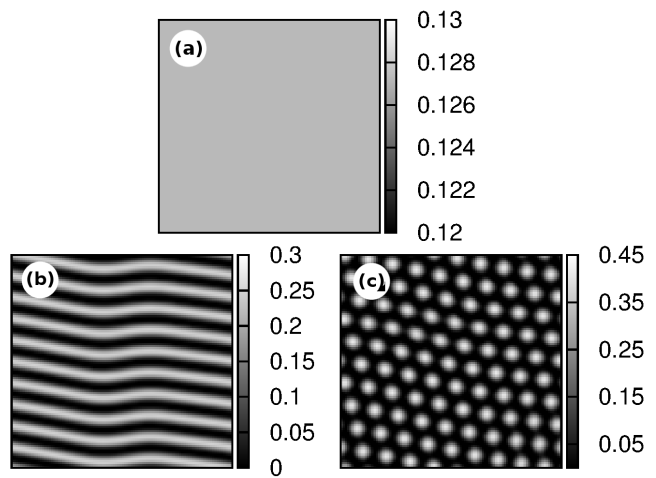}
\caption{Distribution of vegetation produced by the model with a linear probability $h$ for
different values of the competition parameter. $\delta=0.7$ (Left), $\delta=0.8$ (Center), $\delta=0.9$ (Right).
$\alpha_{0}=1$, $\beta=0.1$.}
\label{patterns_recta}
\end{figure}

The simplest function $h$ that fulfills the above-mentioned properties is a linear function,
$h(\tilde{\rho},\delta)=\delta\tilde{\rho}$, which limits the values of
the competition parameter to $0<\delta<1$ so that $h<1$.
Then
\begin{equation}
 \rho_{0}=\frac{\beta}{\beta+\alpha_{0}\delta},
\label{eq:ro0}
\end{equation}
while the perturbation growth rate is given by
\begin{equation}
\lambda({\bf k})=-\frac{\beta}{\beta+\alpha_{0}\delta}\left[\beta+\alpha_{0}\delta\hat{G}({\bf k})\right],
\end{equation}
from which we obtain a transition to pattern ($\lambda$ becomes positive) at a competition strength,
\begin{equation}\label{critdeltarecta}
 \delta_{c}=-\frac{\beta}{\alpha_{0}\hat{G}({\bf k_{c}})},
\end{equation}
where $k_{c}$ is the most unstable mode, which yields the most
negative value of $\hat G$ and is the mode with the highest
growth rate. First note that again the Fourier transform of $G$
must take negative values for patterns to form. Also,
$\alpha_0$ and $\beta$  have to be chosen properly to have
$\delta_{c}\leq 1$. In particular, if  we take $\alpha_{0}=1$,
$\beta=0.1$, and  a top-hat kernel of radius $R=8$, we get
$\delta_{c}\approx0.75$. It is important to remark that spatial
structures result when the maximum death rate, i.e., the death
rate in fully vegetated areas, is much higher than the birth
rate $(\alpha_{0}\gg\beta)$. Otherwise the model shows standard
logistic growth despite the nonlocal spatial couplings and the
distribution of vegetation is homogeneous. Figure
\ref{patterns_recta} shows the different spatial distributions
of vegetation in the stationary state. The homogeneous
distribution is stable when $\delta<\delta_{c}$
(\ref{patterns_recta}a), while patterns (stripes and spots)
exist for  $\delta>\delta_{c}$ (\ref{patterns_recta}b) and
(\ref{patterns_recta}c), respectively.

\subsection{Competition through a nonlocal linear death term}
\label{subsec:lineal}

We next study a natural extension of the kernel based model as
presented in Eq.~(\ref{neuralmodel}) and previous studies
\citep{Borgogno2009}, but with purely competitive
interactions. The local density of vegetation changes in time
because of its local dynamics (logistic growth) and the spatial
interactions (competition) with other points in the domain,
\begin{eqnarray}\label{model}
 \frac{\partial \rho(x,t)}{\partial t}&=&D\nabla^2 \rho(x,t)+\rho(x,t)\left[1-\frac{\rho(x,t)}{\kappa}\right] \nonumber \\
 &-&\Omega\int G_a(|x-x'|)\rho(x',t)dx',
\label{eq:neuralnolocal}
\end{eqnarray}
where $\kappa$, is the {\it carrying capacity} and $\Omega$ is
the {\it interaction parameter}. We have added a diffusive term
modeling seed dispesal. Competitive interactions are determined
by considering the strength of the interactions parameter,
$\Omega$, and the kernel function, $G_a$, both always positive.
This description is equivalent to considering a nonlocal
linear death term which arises from competition among plants.
As mentioned in Subsection \ref{subsec:kernel} the density can
take negative values. This is a consequence of the nonlocal
interactions reinforcing the death of vegetation and entering
linearly on the model; these models are, therefore,
mathematically ill-posed. This is a weakness that these models
share with many related kernel-based models (see
Subsection \ref{subsec:kernel}), but which is absent when
nonlocal competition enters nonlinearly. Negative densities
are nonsensical from a biological point of view, so following
\cite{Borgogno2009}, we set $\rho(x,t)=0$ in model
(\ref{eq:neuralnolocal}) when this occurs. The stationary
solutions are $\rho_{0}=0$ (no vegetated state), and a
nontrivial solution
\begin{equation}\label{steadyhomo}
\rho_{0}=(1-\Omega)\kappa,
\end{equation}
that imposes a constraint on the values of $\Omega <1 $.

\begin{figure}[h]
\begin{center}
\includegraphics[width=\columnwidth]{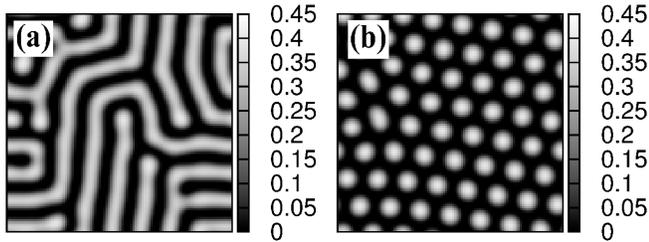}
\caption{Spatial distribution of vegetation for the model described by
Eq.~(\ref{model})
(a) $D=1$, $\Omega=0.7$, and (b) $D=1$, $\Omega=0.9$. $R=8$ in both panels.}
\label{diff}
\end{center}
\end{figure}

 The growth rate of the perturbations is now
\begin{equation}\label{disp}
 \lambda({\bk})=-D|{\bk}|^{2}+1-2\kappa^{-1}\rho_{0}-\Omega \hat{G}({\bk}),
\end{equation}
and using the expression of the homogeneous
steady state, $\rho_{0}$, given by Eq.~(\ref{steadyhomo}), it becomes
\begin{equation}\label{disp2}
 \lambda({\bk})=-D|{\bk}|^{2}-1+2\Omega-\Omega \hat{G}({\bk}).
\end{equation}
There is in this model no restriction on the shape of the Fourier transform of the kernel
for the appearance of patterns (note that $\Omega$ is always lower than $1$). 
We have numerically integrated Eq.
(\ref{eq:neuralnolocal}) in the regime of patterned solutions
and the results are shown in Figures \ref{diff}(a) and
\ref{diff}(b) for two different values of $\Omega$. The same
sequence of spatial structures is obtained as in the other
models.

\section{Derivation of the effective nonlocal description from tree-water dynamics}
\label{sec:derivation}

The models presented in the previous section are all given by a
phenomenological evolution equation for vegetation density.
An open problem is to infer this type of description from a
mechanistic one where the explicit interactive dynamics of
vegetation competing for water is considered.
This would help, in particular, to unveil the origin and properties of the
kernel function.
In this section we present a preliminar (and not fully
satisfactory) attempt to derive the model
presented in \cite{Martinez-Garcia2013}
and discussed in subsection \ref{subsec:birth}
(the derivation corresponding to the nonlocal death model in \ref{subsec:death}
is a straightforward extension of this calculation).

Let us consider a system involving dimensionless vegetation density, $\rho$, and soil-water $w$.
The dynamics is purely local and competitive and takes the form:
\begin{eqnarray}
 \partial_{t}\rho&=&\beta\rho(1-\rho)w-\alpha\rho, \label{biomass}\\
 \partial_{t}w&=&-\mu\rho w-\gamma w+I+D_w\nabla^{2}w, \label{water}
\end{eqnarray}
where the nondimensional positive parameters are:
the seed production rate $\beta$; the vegetation death rate
$\alpha$; the consumption rate of water by vegetation, $\mu$;
the evaporation rate $\gamma$,
and the rainfall, $I$.
Water percolation in the ground is modeled by a diffusion
constant
$D_w$.
Note that this model is a simplified version, which only includes competitive interactions,
of the model presented in \cite{Gilad2004}.

Since the characteristic time scale of the water is much faster than the one of the biomass
we can do an adiabatic
elimination of the variable $w$ (i.e. $\partial_{t}w=0$) so that
\begin{equation}
-\mu\rho w-\gamma w+I+D_w\nabla^{2}w=0,
\end{equation}
and thus
\begin{equation}\label{eqwat}
  \left(D_w\nabla^{2}-\gamma\right)w=\mu\rho w-I,
\end{equation}
whose formal solution can be obtained using Green's functions, $G_d$,
\begin{equation}
 w({\bf x})=\int G_d({\bf x}-{\bf s})(\mu\rho({\bf s})w({\bf s})-I)d{\bf s},
\label{eqw}
\end{equation}
with the boundary conditions $w(x\rightarrow \pm \infty)=0$.
For simplicity we now consider a one-dimensional situation, although
analogous calculations can be done in two dimensions.
The Green's function is the solution
of
 \begin{equation}\label{deltas}
 D_w\partial^{2}_{{\bx}}G_d-\gamma G_d=\delta({\bx}-{\bs}),
\end{equation}
and it is given by
\begin{equation}
 G_d({\bx},{\bs})=-\frac{1}{2}\exp\left(-\sqrt{\frac{\gamma}{D_w}}|{\bx}-{\bs}|\right)
\end{equation}
Taking the nondimensional small number
 $\mu$ as the perturbative parameter, we can further obtain
an approximate expression for $w$ from Eq. (\ref{eqw})
\begin{equation}
 w({\bx})=-IG_{d0}\left[1+\mu\int G_d({\bx}-{\bs})\rho({\bs})d{\bs}+{\cal O}({\mu^2})\right],
\label{eqw2}
\end{equation}
where $G_{d0}=\int G_d({\bx}-{\bs})d{\bs}<0$, since the Green's function is always negative.
 Plugging this in the equation for the biomass density  (\ref{biomass}), we obtain
the closed expression:
\begin{eqnarray} \label{eq}
 &\partial_{t}\rho=\beta\rho(1-\rho)\left\lbrace-IG_{d0}
\left[\mu\int G_d({\bx}-{\bs})\rho({\bs})d{\bs}+1\right]\right\rbrace& \nonumber \\
&-\alpha\rho.&
\end{eqnarray}
Defining the positive nonlocal density
 $\tilde{\rho}=\int G_c({\bx}-{\bs})\rho({\bs})d{\bs}$, where $G_c=-G_d$,
 we can write equation (\ref{eq}) as
\begin{equation} \label{eq2}
 \partial_{t}\rho=\beta \bar r(\tilde{\rho})\rho(1-\rho)-\alpha\rho,
\end{equation}
where we have defined $\bar r(\tilde{\rho})=I|G_{d0}|\left(1-\mu\tilde{\rho}\right)$.

To have a good agreement with the effective nonlocal dynamics
Eq.~(\ref{modelo}), $\bar r>0$ since it
 represents a probability. This is certainly the case
for small $\mu$. Note that some additional conditions on the
normalization of the Green's function have to be imposed to
limit $r$ to values less than $1$. Also $\bar r' (\tilde \rho)=
-I \mu|G_{d0}|$  is always negative, as we expected.

In this particular example  we obtained an exponential kernel
which does not have the finite-range support that would be
associated to the finite root extent. As a consequence, the
Fourier transform of this kernel has no negative components and
then does not lead to pattern formation. The simple modeling of
water dispersion by means of a diffusion constant does not
contain the additional spatial scale associated to root size,
and should be replaced by some mechanism implementing root
effects. On the other side, the finite-range of the kernel is a
sufficient but not a necessary condition for its Fourier
transform to have negative values. It is well-known the
existence of infinite-range kernels whose Fourier transform has
negative values. This is the case of all stretched exponentials
$G(x) \propto \exp ({-|\bx|^p})$ with $p>2$
 \citep{Pigolotti2007}.
Kernels satisfying this are more platykurtic than the Gaussian
function. Work is in progress along this possible line
to obtain pattern-forming kernels.

\section{Conclusions}
\label{sec:conclusions}

In this paper we have reviewed different nonlocal competitive
models of vegetation in water-limited regions where, despite
the absence of facilitative interactions, patterns may still
appear. The obtained sequence of patterns consists on a
stripped structure and spots of vegetation interspersed on the
bare soil forming a hexagonal lattice. We have not been
able to find patterns consisting on spots of bare soil, which
are also typical in models with both competition and
facilitation among plants. In fact, previous works
\citep{Martinez-Garcia2013a} in which the range of the
facilitation was taken to its infinitesimally short-range (i.e.
local) shown these gapped distributions but only in a very
narrow parameter region close to the transition to patterns
line. This is different from standard models with nonlocal
facilitation in which the whole sequence of patterns (gaps,
stripes and spots) appears in a wider parameter's
interval. This may suggest that facilitative interactions,
although not indispensable for the formation of patterns, could
be important in order to promote some of the
structures that have been reported in field observations.
We note in this context that a careful study of the
bifurcation sequences in local vegetation models reveals that
the standard sequence is not fully robust and depends on
nonlinear details of particular models \citep{Gowda2014}.

From a mathematical point of view, nonlocality enters through
an {\it influence} function that determines the number of
plants competing within a range with any given plant. A
first-order approximation of this distance can be given by
(twice) the typical length of the roots, but field measurements
are needed in order to determine the the range over
which individuals of a given plant species can influence their
neighbors. A necessary condition for pattern transitions, 
for the models under study
where the nonlocality is in the nonlinear term,
is the existence of
negative values of the Fourier transform of the influence
function, which always happens, among other situations, for kernel functions
with finite range.

From a biological point of view, competitive interactions alone
may give rise to spatial structures because of the development
of spatial regions (typically located between maxima of the
plant density) where competition is stronger preventing the
growth of more vegetation \citep{Martinez-Garcia2013}.

An unfortunate consequence of the universal character of these
models is that the information it is possible to gain on the
underlying biophysical mechanisms operating in the system just
by studying the spatial distribution of the vegetation is
limited. Many different mechanisms lead to the same patterns.
Although patterns are universal, models should be specific to
each system. This emphasizes the importance that empirical
studies have in developing reasonable models of the behavior of
different systems. Field work may help theoretical efforts by
placing biologically reasonable bounds on the shape and extent
of the kernel functions used in the models, and also by
approximations to the probability of overcoming
competition, $r(\tilde{\rho},\delta)$.

It is important to note that the type of nonlocal models
presented may have localized solutions. This has been studied,
in a different context \citep{Paulau2013}, for a model that
reduces to Eq.~(\ref{sav1}) when the kernels enhance
selfinteractions, i.e., they are of the type $G(x)= F(x)+ a
\delta (x)$ \citep{Hernandez-Garcia2009}.
In plant ecology, mathematical approaches where the interactions among plants
depend on the local biomass density show localized structures as a consequence
of the bistable behavior between the desert state ($\rho_{0}=0$) and the spatially extended
solutions \citep{Lejeune2002}. This result also extends to nonlocal models either
considering the interplay between water and vegetation 
dynamics \citep{Meron2007} or, in
more recent studies using effective equations 
for the vegetation density \citep{fernandez2013strong}.
In this latter case, the authors
explain the formation of fairy circles
(localized barren patches of vegetation) as localized solutions
of spatially non-local models.

Finally, with this work, we aimed to show that, under
certain conditions, nonlocal competition alone may be
responsible for the formation of patterns in semiariad systems.
More interestingly, spatially regular distribution of vegetation appear
regardless of how competitive interactions are
introduced in the different modeling approaches. Certainly, while it may not
be possible to unambiguously identify the model that generates
an observed pattern, the study of the minimal mechanisms giving
rise to pattern formation limits the set of candidate models
(and biological mechanisms) that need to be considered. We hope
that our results shed light on the task of understanding the
fundamental mechanisms -and the possible absence of
facilitation- that could be at the origin of pattern formation
in semiarid systems.

\section*{Acknowledgments}

R.M-G. is supported by the JAEPredoc program of CSIC.
R.M-G., C.L. and E.H-G acknowledge support from FEDER and MINECO
(Spain) through Grants No. FIS2012-30634 INTENSE@COSYP and CTM2012-39025-C02-01 ESCOLA.
C.L. dedicates this work to the memory of his father.


\end{document}